\documentclass[fleqn,usenatbib]{mnras}

\usepackage{newtxtext,newtxmath}

\usepackage[T1]{fontenc}

\DeclareRobustCommand{\VAN}[3]{#2}
\let\VANthebibliography\thebibliography
\def\thebibliography{\DeclareRobustCommand{\VAN}[3]{##3}\VANthebibliography}

\usepackage{graphicx}	
\usepackage{amsmath}	
\usepackage{array}      

\newcolumntype{L}[1]{>{\raggedright\let\newline\\\arraybackslash\hspace{0pt}}m{#1}}
\newcolumntype{C}[1]{>{\centering\let\newline\\\arraybackslash\hspace{0pt}}m{#1}}
\newcolumntype{R}[1]{>{\raggedleft\let\newline\\\arraybackslash\hspace{0pt}}m{#1}}

\title[ETV - the case of V471~Tauri]{Variability of eclipse timing:the case of V471~Tauri}

\author[E. Kundra et al.]{
Emil Kundra,$^{1}$\thanks{E-mail: ekundra@astro.sk (EK)}
\v{L}ubom\'{i}r Hamb\'{a}lek,$^{1}$
Siegfried Vanaverbeke,$^{2,3}$
Pavol Dubovsk\'{y},$^{4}$
Ludwig Logie,$^{2,3}$
\newauthor
Steve Rau,$^{2,3}$
and Franky Dubois$^{2,3}$
\\
$^{1}$Astronomical Institute of the Slovak Academy of Sciences, 059 60 Tatransk\'{a} Lomnica, The Slovak Republic \\
$^{2}$Astrolab IRIS, Verbrandemolenstraat, Ypres, Belgium \\
$^{3}$Vereniging voor Sterrenkunde, Werkgroep Veranderlijke Sterren, Belgium \\
$^{4}$Vihorlat Observatory, Mierov\'{a} 4, 066 01 Humenn\'{e}, The Slovak Republic
}

\date{Accepted XXX. Received YYY; in original form ZZZ}

\pubyear{2021}

\begin{document}
\label{firstpage}
\pagerange{\pageref{firstpage}--\pageref{lastpage}}
\maketitle

\begin{abstract}
The post-common envelope binary V471~Tauri has been an object of interest for decades. V471~Tau shows various phenomena due to its evolutionary state and unique properties, e.g. its magnetic accretion and eclipse timing variation (ETV). Previous authors explained the ETVs by different, sometimes contradictory theories. In this paper, we present and analyse the variability of the eclipse timing of this star. We observed V471~Tauri over the last ten years and covered the second cycle of its 
period variation. Based on our analysis of the presented data, we assess the possible existence of a brown dwarf in this system and derive its orbital parameters. We compare the results of our dynamical modelling to the solution predicted by Applegate-mechanism theories, which have been developed in recent studies. We found that the observed ETV cannot be explained only by the presence of additional components to the binary.
\end{abstract}




\section{Introduction}
\label{sec:intro}

V471~Tau (BD+16~516) is a detached binary that was first studied by \cite{1970PASP...82..699N}. The system consists of a white dwarf and a cool main sequence star of spectral type K2~V with significant chromospheric activity \citep{1984AJ.....89.1252G}. V471~Tau is a member of the Hyades cluster \citep{2001A&A...367..111D}. The binary probably underwent a common envelope evolutionary phase and is therefore thought to be a post-common envelope binary (PCEB) and a pre-cataclysmic variable star \citep{1976IAUS...73...75P, 1985ASSL..113...15B, 2003ApJ...594L..55D}. The precursors of PCEBs are binaries with two unequal mass components. When the more massive star evolves from the main sequence, it becomes a giant star and fills its Roche lobe. The main sequence star cannot incorporate all of the mass transferred from the giant star; therefore, a common envelope forms around the binary. The friction between the gas in the envelope and the companion of the giant results in a spiral-in phase of evolution, in which the companion star moves towards the core of the evolved star, releasing energy that expels the common envelope and finally results in a white dwarf+red dwarf system. The new orbital period is shorter than the original. Magnetic fields and radiation pressure also contribute to the expulsion of the common envelope.

V471~Tau is a white dwarf+red dwarf binary, with an orbital-plane inclination of $77~\mathrm{deg}$ to an Earth-bound observer; hence it shows total eclipses of the white dwarf with an orbital period of $P_{\rm{orb}}\approx\,0.52118357\,\rm{days}$ \citep{2011Ap&SS.331..121K}, or $\approx12.51\,\rm{hours}$. 
The duration of the eclipse is 49~minutes. 
The ingress and egress times take only 55~seconds. The combination of the almost half-day orbital period, the quick decrease and increase phase in its light curve, the long duration of the minimum and its shape, as well as the visibility of V471~Tau on the sky for ground-based observatories limit the number of observed eclipses. Until 2010, only approximately 200~times of minima have been published after 40~years of observation. In this work we present 32 additional times of minima observed in the last ten years. We use the combined dataset to analyse the possible cause of the observed ETV.  Firstly, we derive the orbital parameters of additional components to the binary assuming the ETV is caused by a system of previously undetected substellar companions. Secondly, we explore the energetic feasibility of a magnetic cycle in the secondary star as a cause of the observed ETV via the Applegate mechanism.

In the following section, we review the previous papers which studied the ETV in V471~Tau extensively. Then we describe our data acquisition and reduction, and list the newly presented times of eclipse in section~\ref{sec:extension}. Next, we explain the observations by the dynamical effect of additional components in the system as well as the Applegate mechanism in sections~\ref{subsec:thirdcomp} and~\ref{subsec:applegatemagneto}, respectively. We compare and discuss the models in section~\ref{sec:discussion} and finally present our conclusions in section~\ref{sec:conclusion}.

\section{On the history of eclipse timing variations in V471~Tau}
\label{sec:history}
From the beginning of the study of V471~Tau, it was clear that the observed times of eclipse do not follow a linear ephemeris. The topic of this work is to add to the discussion on the magnitude of ETV caused by different proposed physical effects, discern the differences in the data available to previous authors and to distinguish the mechanism which fits the observables.

The results presented in previous papers were based on the data available at the time of publication and thus depend on a partly covered cycle of the period variations. \citet{2007AJ....134.1206K} discuss three possible interpretations of the changes seen in the observed minus calculated diagram (O-C), including the presence of a third body (see Fig.~6 in \citet{2007AJ....134.1206K}), apsidal motion, and sudden period changes. They concluded that the predicted downward bend in the O-C had not yet happened. 
They explained the observed shape of the O-C curve by apsidal motion which requires only a very small non-zero eccentricity ($e=0.0121$) of the binary orbit.
A sudden period change may explain the O-C in the case of mass transfer or mass loss $\approx~3.8\times\,10^{-7}\,\rm{M_{\odot}\,year^{-1}}$ \citep{2001icbs.book.....H}. This mass loss rate appears to be high and unlikely for the red dwarf in V471~Tau, as it would result in other detectable phenomena. The observed period increase and decrease instead suggest an explanation involving additional complexity of the system \citep{2007AJ....134.1206K}.

\citet{2011Ap&SS.331..121K} found the signs of a long-anticipated downward bend in the O-C. The data were not consistent with some predictions reported in previous publications \citep{2001ApJ...546L..43G, 2007AJ....134.1206K}. They recalculated the parameters of a possible third body in the system. After removing the third body signal, the remaining O-C residuals showed variations which were hypothesized to be connected to the presence of a fourth companion. Another proposed explanation of the residuals was the Applegate mechanism \citep{1992ApJ...385..621A}, which is related to the magnetic activity cycle of the (K2~V) red dwarf companion producing small cyclic changes in the radius of the star.

A cyclic fluctuation in the timing residuals of 5 or 5.5~years was reported by \citet{1994A&A...281..811I, 2005MNRAS.360.1077I}, and \citet{2007AJ....134.1206K} mentioned a 10-year cycle. \citet{2015ApJ...810..157V} found 13, 9, and 5~yr variability cycles, which they fitted with a sinusoid of amplitude 7, 5.7, and 6 seconds, respectively. \citet{2018RNAAS...2..179M} reported the amplitude of the timing residuals fluctuations at 40~seconds.

\citet{2015ApJ...800L..24H} attempted to observe the brown dwarf companion to the V471~Tau binary using the VLT SPHERE instrument by directly imaging the system in the H-band. They took all of the available data and added their own four times of eclipse, measured by ULTRACAM, and calculated the orbital parameters in a single companion and a two companions scenario (including the variations in residuals). Based on the brown dwarf models of \citet{2012RSPTA.370.2765A},  
\citet{2015ApJ...800L..24H} calculated the contrast of the brown dwarf for the H band SPHERE IRDIS instrument as $\delta m_{H}\sim9.2$. Their results were calculated based on a mass for the brown dwarf of $0.044~\rm{M_{\odot}}$ and an age of the Hyades cluster of 625~Myr. 
They expected the brown dwarf signal to be detectable at an angular distance of 260~mas but did not find it in the data.

\citet{2015ApJ...810..157V}, in their multi-data type analysis of this system, calculated the value of the binary period increase $dP/dt$ as $\mathrm{0.286 \times 10^{-10}}$. 
They explained the cause of the slow period increase as the result of a transfer of angular momentum from the white dwarf spin to the binary via magnetic coupling. The white dwarf is in fast rotation with a period of $9.25$~minutes \citep{10.1046/j.1365-8711.1998.01607.x}. \citet{2015ApJ...810..157V} also obtained different values for the detection limit of SPHERE used by \citet{2015ApJ...800L..24H} and noticed features in the image, explaining why \citet{2015ApJ...800L..24H} could not see the brown dwarf companion to the binary. They proposed continued imaging observations and further eclipse timing observations over several decades. According to \citet{2015ApJ...810..157V}, the best time for imaging the brown dwarf was in the time frame 2018-2019, when the O-C was at \emph{zero} level, which corresponds to the maximal distance of the companion from the binary as seen by an observer on Earth.

\citet{2017ASPC..509..571V} discussed the observations by EXOSAT X-ray, XCOV-II Whole Earth Telescope (WET) U-band, and various observations in U-band photometry during the years 1985-1992.
They studied the relation between the spin period of the white dwarf and the orbital period of the binary. They constructed the spin-period O-C diagram and found that the O-C variations related to the white dwarf spin do not match those of the eclipse times. They inferred that the barycenter of the binary does not wobble with the amplitude required to account for the eclipse timing variation and ruled out the third body interpretation. \citet{2017ASPC..509..571V} proposed the Applegate mechanism instead \citep{1992ApJ...385..621A}, but they stated that observational signatures of the Applegate effect still do not exist in the data. However, their data covered only a relatively short (1985-1992) baseline of the $\approx\,45\,\rm{year}$ binary period and had substantial uncertainties.

The next paper about the companion to the V471~Tau binary was published by \citet{2018RNAAS...2..179M}. They analysed 45~years of eclipse timings, including eclipse timings generated from 60~days of short-cadence ($\approx60\,\rm{s}$) $K2$ mission photometry. They reported an ETV caused by a third body with mass $M_{\rm{3}}>~0.043~\rm{M_{\odot}}$ and semi-major axis $13.3\,\rm{AU}$, with an orbital period of $35\,\rm{years}$ and a semi-amplitude of $150\,\rm{seconds}$. Small oscillations with a semi-amplitude of 20~seconds were left in the residuals after the signal caused by the third body had been removed. They connected this residual signal to an additional 4th~body or an Applegate-type mechanism \citep{1992ApJ...385..621A}. They also explained the failure to detect the brown dwarf by \citet{2015ApJ...800L..24H} suggesting that VLT-SPHERE could miss the brown dwarf because V471~Tau is a PCEB system and any hosted body would have been subjected to strong mass outflows, possibly drastically altering its physical properties.

Along with the observations of V471~Tau, theoretical studies of the Applegate mechanism \citep{1992ApJ...385..621A} developed significantly. The Applegate mechanism explains the orbital period modulation of $\Delta P/P\sim\mathrm{10^{-5}}$ as a consequence of gravitational coupling of the orbit to variations in the shape of a magnetically active star in the system. The changing distribution of angular momentum can produce deformation of the active star during its activity cycle. A subsurface magnetic field of several kilogauss can provide the torque needed to redistribute the angular momentum.

\citet{2005MNRAS.364..238L} introduced a general model for the angular momentum transfer within a turbulent stellar convection zone and discussed the model within the context of the Applegate hypothesis in the case of a magnetically active RS~CVn binary. They found that an angular velocity change of 10 per cent of the unperturbed angular velocity at the base of the stellar convection zone is needed and  concluded that such a change is not compatible with the observations because it would produce an energy dissipation rate much larger than the typical luminosities of the active components of RS~CVn systems. \citet{2006MNRAS.369.1773L} showed, based on his previous work, that consideration of the Applegate hypothesis leads to the same energy balance problem in the case of the secondary components of cataclysmic variables and related systems. They noted that the anisotropic Lorentz force due to an internal magnetic field may produce the observed orbital period modulation. 

\citet{2020MNRAS.491.1820L} continued his work and introduced a new model to explain the modulation of the orbital period observed in close stellar binary systems based on an angular momentum exchange between the spin of the active component and the orbital motion. The spin-orbit coupling can be produced by a non-axisymmetric component of the gravitational quadrupole moment of the active star due to a persistent non-axisymmetric internal magnetic field. 
Based on the modulation period published by \citet{2015ApJ...810..157V}, they estimated the required energy to be $3\times10^{-3}$ and $\approx7\times10^{-4}$ of the energy produced by the secondary star of V471~Tau for their circulation and libration models, respectively. 

\citet{2018A&A...615A..81N} significantly contributed to the magneto-hydrodynamic models of stars. They explored the energetic feasibility of driving eclipsing time variations by the Applegate mechanism in the formulation presented by \citet{2016A&A...587A..34V}. For a sample of the PCEB systems, they used different radial density profiles to investigate the radii at which the predicted activity cycle matches the observed modulation period. They found that the radius where the activity cycle matches the period modulation depends on the level of activity of the star, expressed as a parameter $\gamma$, and assumed that the dynamo is driven in the outer layers at approximately the relative core-shell transition radius $R_{d}/R$. For V471~Tau, the calculated relative core-shell transition radii $R_{d}/R$ are 0.78 and 0.87 for values of the parameter $\gamma$ set to 0.86 and 1.0, respectively. They investigated the energetic feasibility of the Applegate mechanism in terms of the fraction of the available energy $\Delta E/E$  within the companion star which is required to drive the observed magnetic cycle. They found that for different $R_{d}/R$, there is a wide minimum of $\Delta E/E$ near $R_{\rm{core}}/R\sim0.6$. They calculated $\Delta E/E$ for $R_{d}/R=0.75$ and 0.85. For V471~Tau, they reported the corresponding values for $\Delta E/E$ as 0.061 and 0.036, respectively. They concluded that, for the secondary in this binary, a Sun-like star with mass $\approx0.93 M_{\odot}$ rotating much faster than the Sun, one may expect that a strong magnetic field, due to the rapid rotation, can indeed trigger the Applegate mechanism. They concluded that the energy produced by the secondary star in V471~Tau is sufficient to drive the change in the gravitational quadrupole moment and generate the orbital period changes implied by the observed ETVs.


\citet{2018A&A...615A..81N} created the Applegate online tool\footnote{Applegate calculator:\newline\url{http://theory-starformation-group.cl/applegate}} which computes the energetic feasibility of driving eclipsing time variations via the Applegate mechanism. The input parameters in the calculations are the binary period, binary semi-major axis, modulation period, observed period change $\Delta P/P$, mass, radius, luminosity, temperature of the magnetically active star, and the parameters $k_{1}, k_{2}$. The parameters $k_{1}$ and $k_{2}$ depend on two mean densities inside the star: $\rho_{in}$ from the bottom to $R_{d}$, and $\rho_{out}$ from $R_{d}$ to $R_{star}$. Their model assumes an interchange of angular momentum between these two zones.
\citet{2020MNRAS.491.1043N} continued the theoretical study of the Applegate mechanism and presented two compressible non-ideal magneto hydrodynamical simulations of the magnetic dynamo in a solar mass star. They ran their simulations for two different rotation rates, one of them with three times the solar rotation rate and the other with 20~times the solar rotation rate. They found that both the magnetic field and the stellar quadrupole moment change in a quasi-periodic manner. For the rapid rotator, the behaviour of the magnetic field, as well as the quadrupole moment changes, became considerably more complex due to the less coherent dynamo solution. The resulting O-C variations of the simulation were of the order of 0.35~s for the slow rotator model and 23~s for the fast rotator model. These values are less than the 151 and 20~s cycles reported before in V471~Tau(\citep{2018RNAAS...2..179M}). They explained that their simulations may not have captured all of the relevant effects because they did not take into account the centrifugal force and self-gravity. Also, the rotation of V471~Tau is still a factor 2.5~faster than their rapid rotator model. \citet{2020MNRAS.491.1043N} found strong evidence that magnetic effects can indeed produce eclipsing time variations. Recently, \citet{2022arXiv220503163N} have improved their models and included the centrifugal force in the MHD simulations. They found variations of the stellar quadrupole moment irrespective of the presence of the centrifugal force. They stated that the original models proposed by \citet{1992ApJ...385..621A} have difficulties explaining the observed magnitude of the variations, and the centrifugal force is unlikely to be the main driver of the variations.

Spectropolarimetric observations of the K2~dwarf star in V471~Tau were reported in the study by \citet{2021MNRAS.504.1969Z}. They found differential rotation fluctuations that do not favour an Applegate mechanism operating in the V471~Tau system, at least in its standard form. Based on the work of \citet{2018A&A...620A..42V}, they were unable to explain the observed ETVs. 
\citet{2022MNRAS.513.2893Z} analysed the spectropolarimetric data of the binary system V471~Tau
collected from December 20, 2014 to January 12, 2015, with ESPaDOnS and compared the results with their
previous study on data from 2004 - 2005 \citep{2021MNRAS.504.1969Z}. \citet{2021A&A...650A.158K}
presented the results of the X-ray luminosity of the system. They reported that the long-term evolution
of X-ray luminosity reveals a possible activity cycle length of $\approx12.7~\mathrm{yr}$, traces of
which were also discovered in spectra of the $H_{\alpha}$ line strength. 
Also, \citet{2022RNAAS...6...94C} analysed X-ray observation from different satellite missions since 1980. Their period searches yielded tentative periods of $\sim~3.9$, $\sim5.5$ yr and $\sim11.9$ yr. They didn't found an evidence for a $\sim36$ yr magnetic-dynamos activity cycle in the X-rays.
\citet{2022MNRAS.513.2893Z} found that the average magnetic field strength increased by approximately 2.2 times from the two first epochs (2004-2005, at an activity minimum) to the 2014-2015 epoch (at the activity maximum). They did not find such a modulation in the brightness maps, which display a spot coverage of 14 per cent, 17 per cent, and 18 per cent in 2004.9, 2005.9, and 2014.9/2015.1, respectively. Their result emphasizes that the spot coverage may not always be an appropriate observable with which to study activity cycles in very active rapidly rotating stars.

V471~Tau is not the only binary showing eclipsing time variations. \citet{2016MNRAS.460.3873B} presented an eclipsing time monitoring program aimed at revealing the extent and amplitude of eclipsing time variations throughout the class of binaries with white dwarf components. Their work will enable a systematic search for correlations between the level of eclipsing time variability and the physical characteristics of the systems, such as the secondary star's spectral type. The observing campaign includes V471~Tauri and 66 other close binaries, comprising detached and semi-detached systems with M-dwarfs, K-dwarfs, brown dwarfs and white dwarf secondary stars. Based on more than 650~white dwarf eclipses observed at the beginning of this campaign, \citet{2016MNRAS.460.3873B} showed that all binaries with observational baselines exceeding 10~yr. and secondary stars of spectral type K2-M5.5  show variations in the eclipse arrival times that, in most cases, amount to several minutes. \citet{2016MNRAS.460.3873B} also found that binaries with a late spectral type ($>$M6) red dwarf, brown dwarf or white dwarf secondary star appear to show no orbital period variations on baselines shorter than 10~years. Their results agree with the so-called Applegate mechanism, which proposes that magnetic cycles in the secondary stars can drive variability in the binary orbits. \citet{2016MNRAS.460.3873B} concluded that a complex eclipsing time variation exists in binaries with significant O-C variations such as V471~Tau, HU~Aqr, and QS~Vir.

A recent paper by \citet{2022arXiv220606919P} discussed whether the ETV is a reliable indicator of circumbinary companions in post-common envelope binaries. \citet{2022arXiv220606919P} presented 163 new times of minima of seven PCEBs (HW~Vir, NY~Vir, V470~Cam, NSVS~14256825, NN~Ser, R~Cae and DE~CVn) with confirmed exoplanets listed in the NASA Exoplanet archive \footnote{\raggedright NASA Exoplanet Archive: \url{https://exoplanetarchive.ipac.caltech.edu}} or in the Extrasolar Planets Encyclopedia \footnote{\raggedright Extrasolar Planets Encyclopedia: \url{https://exoplanet.eu}} and tested the existing models. The authors found that more than 30~models of circumbinary companions for the reported PCEBs have failed to predict the data. More observations are needed to confirm the recent model of NY~Vir. They could not explain the observed quasi-cyclical ETVs of the studied PCEBs on the basis of angular momentum loss although apsidal motion and the Applegate mechanisms may contribute to these variations. A magnetic mechanism has a significant influence only for two of their PCEBs, RR~Cae and DE~CVn. However, the presence of a circumbinary object is masked by the magnetic effects associated with the secondary component in PCEBs. Their main conclusion is that increasing observational timelines to at least two circumbinary orbits, considering less precise data, could lead to a better prediction of future eclipse times. 

To distinguish between Applegate modulation and period variation caused by other bodies, long-term measurements are required to determine the correlations between activity indicators, luminosity variations, and the binary period modulation \citep{2018A&A...620A..42V, 2012AN....333..754P, 1992ApJ...385..621A}.
V471~Tau is an important astrophysical laboratory for deciphering binary star evolution, particularly with regard to the PCEB and pre-cataclysmic variable star systems. 


\section{Extending the existing data-set}
\label{sec:extension}
\subsection{Observation and data reduction}
\label{subsec:data}

We obtained the photometric data used in this study at the Observatory of the Astronomical Institute of the Slovak Academy of Sciences in Star\'{a} Lesn\'{a} (AISAS), at the Kolonica Observatory operated by Vihorlat Observatory in Humenn\'{e}, and AstroLAB IRIS observatory in Ypres Belgium. All observations were taken in the Johnson \textit{B} filter. In this filter, the depth of the primary eclipse is 0.2~mag. 

The observations at the AISAS observatories and the Kolonica Observatory are described in \citet{2011CoSka..41...39H}. The AstroLAB IRIS observatory in Zillebeke, Belgium used their 684 mm aperture Keller F4.1 Newtonian New Multi-Purpose Telescope (NMPT). The CCD detector assembly of this instrument is a Santa Barbara Instrument Group (SBIG) STL 6303E operating at $ -20\,^{\circ}\rm{C}$. A 4-inch Wynne corrector feeds the CCD at a final focal ratio of 4.39, providing a nominal field of view of $20\times 30\,\rm{arcmin}$. The $9\,{\mu}m$ physical pixels project to $0.62$\,arcsec per pixel and are read out binned to 3x3 pixels, i.e. $1.86\,\rm{\mu}$m per combined pixel. The \textit{B}, \textit{V}, and \textit{R} filters are from Astrodon Photometrics and have been shown to reproduce the Johnson-Cousins system closely\footnote{\raggedright Astrodon Photometrics Test Summary: \url{https://astrodon.com/wp-content/uploads/2018/05/Astrodon-Photometrics-UVBRI-Filters_shendentestsummary.pdf}}. An overview of the instruments is provided in Table~\ref{tab:KSG1G2}.

\begin{table}
\caption{Basic data for V471~Tau and comparison stars used in photometry}
\centering
\begin{tabular}{cc}
\hline
\multicolumn{2}{c}{V471~Tau and comparisons} \\
\hline
V471~Tau& GSC~01252-00212, BD~+16~516 \\
$\rm{RA_{2000}}$ [h m s] & $\;03\,50\,24.97$ \\
$\rm{DEC_{2000}}$ [$^{\circ}\, ^{\prime}\, ^{\prime\prime}$] &$+17\,14\,47.42$\\
$d$ [pc]& $48$\\
$B$ [mag] & $10.24$ \\
$V$ [mag] & $9.48$ \\
spectral type & WD + K2~V\\
orbital period [h]& $\sim\,12.5$\\
\multicolumn{2}{c}{comparison star BD +16~515} \\
$B$ [mag]& 10.73 \\
$V$ [mag]& 9.46 \\
\multicolumn{2}{c}{check star USNO-A2.0~1050-01038658} \\
$B$ [mag]&12.6 \\
$V$ [mag] &11.7 \\
\hline
\end{tabular}
\label{tab:v471tau_katalog}
\end{table}

We focused our observations on the rapid ingress and egress phase of the total eclipse, which take only 55~seconds. In order to improve the time resolution of the data, we used short exposure times and binned the image frames to 2x2 pixels for AISAS and KO and 3x3 pixels for the NMPT, respectively. The exposure times were set to 5, 7, 10 and 15~seconds, respectively, depending on the instrument used and the weather conditions. The photometric precision was approximately 0.01 - 0.015~mag, larger for shorter exposures. In our FOV ($\approx\,15 \times 15\,\rm{arcmin}$ for instruments at AISAS), there are only three stars bright enough for  photometry with short exposure times, see Table~\ref{tab:v471tau_katalog}. We used the same comparison and check stars for all photometric observations.

\begin{table*}
\small
\centering
\caption{Geographical location and technical specifications of the telescopes and detectors
placed at the AISAS Star\'{a} Lesn\'{a} G1 and G2, Kolonica Observatory, both in Slovakia and Astrolab IRIS NMPT-68 in Belgium. The observations were taken only in the Johnson B band.}
\begin{tabular}{cccccc}
\hline
Observatory & \multicolumn{2}{c}{Star\'a Lesn\'a, G1} & Star\'a Lesn\'a, G2 & Kolonica saddle & Astrolab IRIS\\
\hline
obs. abbreviation & G1S & G1F & G2F & KOF & NMPT\\
LONG & \multicolumn{2}{c}{$20^{\circ}\,17^{\prime}\,21^{\prime\prime}\;\mathrm{E}$} & $20^{\circ}\,17^{\prime}\,28^{\prime\prime}\;\mathrm{E}$ & $22^{\circ}\,16^{\prime}\,26^{\prime\prime}\;\mathrm{E}$ & $02^{\circ}\,54^{\prime}\,45^{\prime\prime}\;\mathrm{E}$\\
LAT & \multicolumn{2}{c}{$49^{\circ}\,09^{\prime}\,06^{\prime\prime}\;\mathrm{N}$} & $49^{\circ}\,09^{\prime}\,10^{\prime\prime}\;\mathrm{N}$ & $48^{\circ}\,56^{\prime}\,06^{\prime\prime}\;\mathrm{N}$ & $50^{\circ}\,49^{\prime}\,02^{\prime\prime}\;\mathrm{N}$ \\
altitude [m]& \multicolumn{2}{c}{$785$}& $785$ & $460$ & \\
 &\multicolumn{2}{c}{before and after May 14th 2012} & & & \\
manufacturer & J. Drbohlav & Zeiss & Zeiss & AO of National & \\
 & Czech Republic &Jena Germany &Jena Germany & University in Odessa & \\
type & Newton & Cassegrain & Cassegrain & modified optical sys. & \\
 & reflector & reflector & reflector & Argunov - Faschevsky &\\
diameter & 0.5~m & 0.6~m & 0.6~m & 1.0~m & 0.68~m\\
focal length & 2.5~m & 7.5~m & 7.5~m & 9.0~m & 3.0~m\\
CCD 2010-2011 & SBIG ST10MXE& & & FLI PL1001E & \\
CCD 2012-2015 & & FLI ML~3041 & MI~G4-9000 & FLI PL1001E & \\
CCD 2016-2020 & & & FLI ML~3041 & & SBIG STL 6303E \\
\hline
\multicolumn{6}{c}{\footnotesize Note: We list only those CCD detectors used for the V471~Tau observations.}
\end{tabular}
\label{tab:KSG1G2}
\end{table*}

The data were reduced in a standard way using the DAOPHOT package in \texttt{IRAF} \footnote{\url{https://github.com/iraf-community/iraf}}, custom-writen scripts and FORTRAN programs created by \citet{Priv_Comm_Pribulla_Phot_Reduction} and the authors of this paper. In the beginning, master darks and flats were produced. Then all object frames were cleaned from bad pixels and photometrically calibrated. All frames were astrometrically solved to define the pixel to WCS1 transformation and finally aperture photometry was performed.

\subsection{Estimating times of eclipse} \label{subsec:times}

\begin{table}
\small
\centering
\caption{The observed times of eclipse of V471~Tau in units of BJD-TDB.}
\label{tab:tbl-minima}
\begin{tabular}{cccc}
\hline
Date & Times of eclipse (BJD-TDB) & Instrument & Exp(s) \\
\hline
 9. Oct 2010 & 2455479.428296 $\pm$ 0.000054 & G1S & 10 \\
 5. Oct 2010 & 2455506.529634 $\pm$ 0.000124 & G1S & 10 \\
17. Dec 2010 & 2455548.224443 $\pm$ 0.000056 & G1S &  5 \\
 6. Nov 2011 & 2455872.400341 $\pm$ 0.000060 & G1S &  5 \\
 8. Nov 2011 & 2455874.485116 $\pm$ 0.000081 & G1S &  5 \\
17. Nov 2011 & 2455883.345325 $\pm$ 0.000042 & G1S &  5 \\
11. Dec 2011 & 2455907.319802 $\pm$ 0.000074 & G1S &  5 \\
22. Feb 2012 & 2455980.285448 $\pm$ 0.000103 & G1S & 10 \\
 5. Mar 2012 & 2455992.272326 $\pm$ 0.000052 & G1S & 10 \\
17. Oct 2012 & 2456218.466374 $\pm$ 0.000063 & G1F &  5 \\
18. Oct 2012 & 2456219.508620 $\pm$ 0.000021 & G1F &  5 \\
20. Nov 2012 & 2456252.343262 $\pm$ 0.000036 & KOF & 10 \\
 5. Sep 2013 & 2456541.600094 $\pm$ 0.000029 & KOF & 10 \\
 8. Oct 2013 & 2456574.434847 $\pm$ 0.000153 & G1F &  5 \\
 8. Oct 2013 & 2456574.434610 $\pm$ 0.000035 & KOF & 10 \\
31. Oct 2013 & 2456597.366697 $\pm$ 0.000034 & KOF & 10 \\
 1. Nov 2013 & 2456598.408895 $\pm$ 0.000071 & G1F &  7 \\
 3. Feb 2014 & 2456692.222190 $\pm$ 0.000027 & G1F &  7 \\
 2. Oct 2014 & 2456933.530012 $\pm$ 0.000031 & G1F &  7 \\
12. Oct 2014 & 2456943.432604 $\pm$ 0.000037 & G1F & 10 \\
25. Oct 2014 & 2456956.462037 $\pm$ 0.000031 & G1F &  7 \\
 3. Nov 2014 & 2456965.322353 $\pm$ 0.000109 & G1F & 10 \\
30. Aug 2015 & 2457265.523874 $\pm$ 0.000027 & G1F &  7 \\
31. Aug 2015 & 2457266.566271 $\pm$ 0.000025 & G1F &  7 \\
12. Sep 2015 & 2457278.553286 $\pm$ 0.000050 & G2F &  7 \\
26. Feb 2019 & 2458541.381024 $\pm$ 0.000034 & NMPT & 15 \\
27. Feb 2019 & 2458542.423155 $\pm$ 0.000050 & NMPT & 15 \\
18. Dec 2019 & 2458836.370788 $\pm$ 0.000030 & G2F &  5 \\
29. Dec 2019 & 2458847.315951 $\pm$ 0.000030 & G2F &  5 \\
21. Jan 2020 & 2458870.248052 $\pm$ 0.000091 & G2F &  5 \\
23. Jan 2020 & 2458872.332922 $\pm$ 0.000091 & G2F &  5 \\
18. Sep 2020 & 2459111.556416 $\pm$ 0.000055 & G2F &  5 \\
\hline
\end{tabular}
\end{table}

We note that the light curve (LC) of V471~Tau changes shape continuously due to spot activity on the surface of the K2~V red dwarf companion. 
We used only a part of the LC before and after an eclipse to calculate the times of mid-eclipse of an eclipse. We corrected the eclipse part of the LC for the local slope created by the photometric wave with the \texttt{EQWREC2} \footnote{\url{https://www.ta3.sk/~budaj/software/}} code to eliminate the influence of  starspots on the LC. The photometric points just before and after the eclipse were fitted with a linear function and the eclipse part of the LC was de-trended. The contact points of the eclipses are not well-defined in our observations because of the limits of the time resolution achievable with the instruments used. The exposure times used are too long to cover precisely the four contact points, but we still have at least three photometric points during the rapid ingress or egress phase. There is no evidence of changes in depth or width of the eclipse in the literature or our data. Therefore, we created a simple model of the eclipse with the \texttt{ROCHE} code \citep{2004ASPC..318..117P} as a fit to the detrended observations taken in subsequent nights. We fitted the data with the modelled LC and estimated the times of mid-eclipses. 

The estimated times of eclipse and their errors are collected in Table~\ref{tab:tbl-minima}. The errors depend on the time resolution of the data, the exposure time used, and the readout time of the CCD camera \citep[see][]{2012AN....333..754P}. We consider them as the minimal error values for the times of eclipse.
The real errors of the quoted eclipse times are a few seconds, and they are well below the changes observed in the O-C diagram.

To investigate an effect which has an amplitude of the order of $\approx~2.3$~minutes, we need to consider second-order effects, one of which is the decelerating rotation of the Earth over a $\approx\,50\,$year-long interval. This accumulates to somewhat over 1~minute during these past decades. Another second-order effect we considered is the movement of the Solar-system barycentre that causes shifts of 4~seconds. In this work, we use BJD-TDB consistently. Using Barycentric Dynamical Time (TDB) takes into account relativistic effects and is a truly uniform time which would be measured on Earth if it were not moving around the Sun or was not accelerated by the gravitational field of the Moon and other celestial bodies. All times of eclipse used in this work were, if necessary, transformed to BJD-TDB \footnote{\url{https://sirrah.troja.mff.cuni.cz/~mary/BarCor.pdf}}. 

\section{Possible explanations for the observed period variations}

Various authors have arrived at considerably different values of the period variations seen in the O-C diagram of V471~Tau; e.g. \citet{1992ApJ...385..621A} reported a period modulation of $\approx20~\rm{years}$. With more eclipse times becoming available, \citet{2018RNAAS...2..179M} found a longer period in the range of 30 to 35 years. Previously published analyses therefore cover no more than 1.5 cycles of the period variation.

Our 32~presented times of eclipse were added to the values published by \citet{1970PASP...82..699N, 1972ApJ...173..653Y, 
1972IBVS..657....1A, 1974A&A....36..459L, 1975PASP...87..461Y, 
1976A&A....46..197C, 
1978IBVS.1444....1O, 1978Ap&SS..57..219I, 1979ApJ...230L.187B, 
1981IBVS.1924....1P, 1982JAVSO..11...57S, 1982IBVS.2189....1P, 
1983ApJ...267..655Y, 
1984IBVS.2573....1I, 1985IBVS.2670....1B, 1987IBVS.3078....1P, 
1987IBVS.3095....1E, 1988AJ.....96..976S, 1989IBVS.3355....1K, 
1989Ap&SS.161..221I, 1992IBVS.3760....1W, 1993Ap&SS.204..297T, 
2001ApJ...546L..43G, 2005MNRAS.360.1077I, 2007JSARA...1...17M, 
2011Ap&SS.331..121K, 2011CoSka..41...39H, 2015ApJ...800L..24H, 
2018RNAAS...2..179M, 2022RNAAS...6...94C}.
All of the previously published times of eclipse used in the presented study are available as supplementary material.

The data presented in this work cover the second cycle of the period modulation allowing us to specify the period more precisely and analyse the physical mechanisms for these variations. In the following sections, we will look at two possible explanations which were extensively discussed in the previous literature on this system. We will consider a possible additional component to the binary and the Applegate mechanism from the perspective of theories and models which were developed in recent years (for an overview see \citet{2018A&A...615A..81N}).

\subsection{A third component solution}
\label{subsec:thirdcomp}

The 32 new times of eclipse reported in this work cover the second cycle of the eclipse timing variation and enhance the time base of the O-C to $\approx\rm{50~years}$. This allows us to investigate the presence of an additional companion to the binary, which should cause strictly periodic variations.

\begin{figure}
\includegraphics[width=\columnwidth]{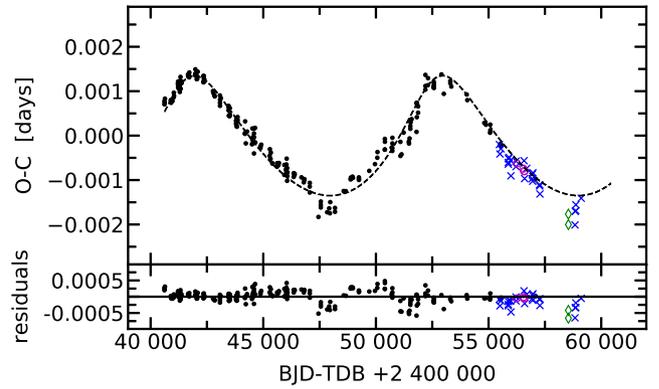}
\caption{The O-C diagram for V471 Tau constructed using the new ephemeris derived in this work, which includes the effect of the third component in the system. The bottom panel shows the residuals remaining after modelling the third component. (dots - previously published, crosses - G1S+G1F+G2F, diamonds - NMPT, circles - KOF data)}
\label{fig:2021OC}
\end{figure}

We applied a non-linear ephemeris to construct the O-C diagram. As a starting point, we used the ephemeris calculated from the "all data" solution by \citet{2015ApJ...810..157V}. Their ephemeris includes the period increase $dP/dt = \mathrm{0.286 \times 10^{-10}}$. We computed a new ephemeris with a fixed value of the period increase of the white dwarf - red dwarf binary using all of the available times of eclipse, see Fig.~\ref{fig:2021OC}. Based on our calculated ephemeris, we modelled the effect of an additional body to the binary using the code \texttt{3T} \citep{2005ASPC..335..103P}. The best-fit solution corresponds to a companion on an eccentric orbit with a period of $\approx\rm{30~years}$. The resulting orbital parameters are collected in Table~\ref{tab:v471tau_3t}. 
Note that we calculated the parameters of the companion concerning its orbit center. The equation used in the code \texttt{3T} includes the term $e\sin\omega_{3}$, and the values in our O-C diagram extend equally above and below zero (see~\citet{1952ApJ...116..211I, 1959AJ.....64..149I}).
 
Using the total mass of the binary $M_{1}\,+\,M_{2}=1.72\,\rm{M_{\odot}}$ derived by \citep{2007AJ....134.1206K} and for typical inclinations of the component orbit $i_{3b}$ in the range $45~\mathrm{deg}\,-\,85~\mathrm{deg}$, the mass of the third component is found to be in the range $M_3=37\,-\,52\,\rm{M_{J}}$. For $i_{3b}=85\,~\mathrm{deg}$, the calculated mass is $0.035\,\rm{M_{\odot}}$; hence, the third companion has a mass consistent with a brown dwarf. Individual fits for different inclinations are displayed in Table~\ref{tab:v471tau_3t-massdistance}.

For the calculation of the position in the sky, we used the code described in \citet{2018A&A...616A..49P}. We adopted the distance to V471~Tau as $49\,\rm{pc}$ \citep{2015ApJ...810..157V}, a value close to the distance of $50\,\rm{pc}$ quoted by \citet{2015ApJ...800L..24H}. The position angle of the ascending node of the third component's orbit is unknown. We therefore used an arbitrarily fixed value of $\pi/2$. The resulting separation between the third component and the binary for the date of the SPHERE VLT observation and the predicted maximal separation are in the range of 210-280 mas (see Table~\ref{tab:v471tau_3t-massdistance}). The residuals left after removing the orbit of the third body are shown in the bottom plot of Figure~\ref{fig:2021OC} and show a wave-like structure. We analysed the timing residuals using the methods implemented in the \texttt{Peranso} software tool \citep{2016AN....337..239P}. The spectral window analyses of the timing residuals (see~Fig.~\ref{fig:2021spectralwindow}) revealed a dominant period of approximately one year which is caused by the time distribution of the data. The location of the observatories in the northern hemisphere and the position of  V471~Tau in the sky affects the possibility of the V471~Tau observations. The date compensated discrete Fourier transform (DCDFT) (see~Fig.~\ref{fig:2021dcdft}) resulted in a wide peak at a period of $\approx4000~\rm{days}$ ($\approx11~\rm{years}$) accompanied by smaller peaks at $\sim1500, 1900$ and $2700~\rm{days}$. 
We didn't find a period supporting an additional companion to the binary, a separate fourth body or a companion to the third component. The wide peak at $\sim$11-year instead favours a quasi-periodic cycle.

\begin{table}
\caption{Physical and geometrical parameters of the third component of the V471~Tau system. $P_{3}$ - orbital period of the third body, $e$ - eccentricity of the third component, $\omega_{3}$ - argument of periastron, $T_{\rm{periastron}}$ - time of periastron passage, $a_{12}\sin(i)$ - projected semi-major axis of the eclipsing pair around the common centre of gravity, 
$T_{0}$ - time of minimum of the close binary selected for zero epoch, $P_{0}$ - period of the close binary, $f(M_{3})$ - mass function of the third component.}
\centering
\begin{tabular}{ccc}
\hline
Parameter & Value & Unit \\
\hline
$P_{3}$ & $11\,016\, \pm\, 350$ &days\\
$e$& $0.393\, \pm\, 0.05$ & \\
$\omega_{3}$&$1.155\, \pm\, 0.1$ &rad\\
$T_\mathrm{periastron}$ &$2\,441\,516\, \pm\, 150 $ & BJD-TDB\\
$a_{12}\sin(i)$&$0.234 \pm 0.01 $& AU\\
$T_{0}$ & $2\,445\,821.897751\pm\,2.1\times 10^{-5}$ & BJD-TDB\\
$P_{0}$& $0.5211833875 \pm\,2.01\times 10^{-9}$& day\\
$dP/dt$&$0.286\times 10^{-10}\pm\,0.011$&\\ 
$f(M_{3})$& $1.4113 \times 10^{-5}\pm\,6.69\times 10^{-7}$ & $\mathrm{M_{\odot}}$ \\
\hline
\multicolumn{3}{L{0.9\columnwidth}}{\footnotesize Note: The parameters are computed with respect to the orbit's centre
(see~\citet{1959AJ.....64..149I}).}
\end{tabular}
\label{tab:v471tau_3t}
\end{table}

\begin{table}
\caption{Values for the mass of the brown dwarf assuming a total mass of $1.72\,\rm{M_{\odot}}$ for the binary and various values of the inclination. The other columns are the separation between the brown dwarf and the binary at the date of the SPHERE VLT measurement ($d_{\rm{SPHERE}}$) and the maximum separation  ($d_{\rm{max}}$).}
\centering
\begin{tabular}{ccccc}
\hline
$i_{\rm{3}}$ & \multicolumn{2}{c}{$M_{\rm{3}}$} & $d_{\rm{SPHERE}}$& $d_{\rm{max}}$ \\
$\rm{\deg}$& $\rm{M_{\odot}}$ & $\rm{M_{J}}$ & mas & mas \\
\hline
$85$ & $0.035$ & $37$ & $210$ & $260$ \\
$75$ & $0.036$ & $38$ & $218$ & $261$ \\
$65$ & $0.039$ & $41$ & $234$ & $264$ \\
$55$ & $0.043$ & $45$ & $253$ & $269$ \\
$45$ & $0.050$ & $52$ & $273$ & $278$ \\
\hline
\end{tabular}
\label{tab:v471tau_3t-massdistance}
\end{table}

\begin{figure}
\includegraphics[width=\columnwidth]{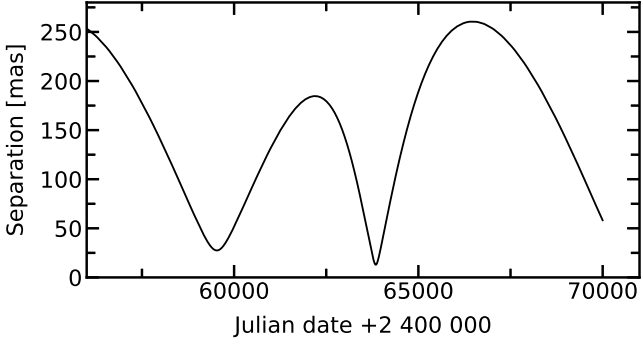}
\caption{The separation of the third companion from the white dwarf - red dwarf binary for an orbit with an inclination of $85~\mathrm{deg}$. The values of the separation are calculated using the model described in section \ref{subsec:thirdcomp}. 
}
\label{fig:separation}
\end{figure}

\begin{figure}
\includegraphics[width=\columnwidth]{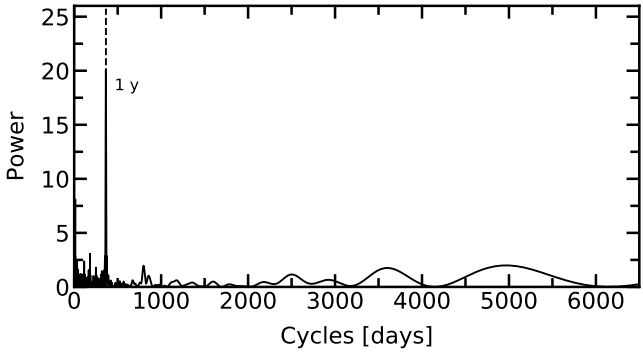}
\caption{Spectral window of the timing residuals based on an extended dataset. The one-year period which is caused by the time distribution of the data is highlighted with a vertical dashed line.}
\label{fig:2021spectralwindow}
\end{figure}

\begin{figure}
\includegraphics[width=\columnwidth]{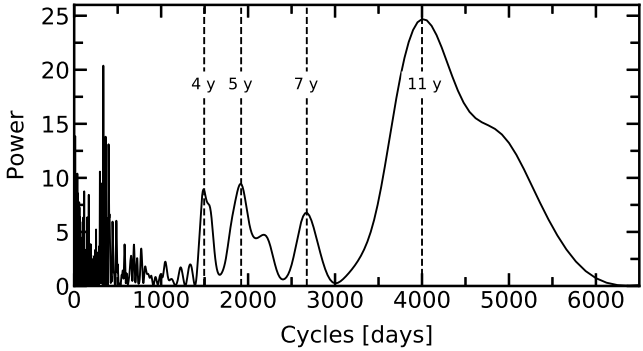}
\caption{DCDFT analysis periodogram of the residuals. The significant cycles of approximately 4, 5, 7, and 11 years are highlighted with vertical dashed lines.}
\label{fig:2021dcdft}
\end{figure}
\subsection{Applegate mechanism}
\label{subsec:applegatemagneto}

Let us now investigate the explanation of the observed O-C variation by the Applegate mechanism and the magneto-hydrodynamic models developed in a number of recent studies (e.g. \citet{2018A&A...615A..81N, 2020MNRAS.491.1820L} and references therein).

We consider two possibilities: i) the Applegate mechanism is responsible for the period modulation and ii) the larger amplitude cycle ($\approx300~\rm{seconds}$) is caused by the presence of a third body, while the effect of the Applegate mechanism is seen only in the variations of the residuals after subtracting the model for the orbit of the third component derived in the previous section.

We used the online Applegate calculator to investigate the energy required to power a magnetic cycle in the companion star of V471 Tau. We used the previously determined period of the modulation, and the O-C amplitude computed from the orbit of the third component modelled in the previous section, as input values for the calculator. From the period modulation of $P_{3}\approx30~\rm{years}$ and the $230~\rm{second}$ amplitude of the brown dwarf O-C, we calculated the observed relative change of the orbital period during one cycle of the binary to be $\Delta P/P=1.5\times10^{-6}$ (for more details see Equation~38 in \citet{1992ApJ...385..621A}). 
We used this parameter, together with the period of the binary $P_{0}\sim 0.52118~\rm{day}$, as additional inputs to the Applegate calculator, while we used for the parameters of the magnetically active star (mass $0.93~\rm{M_{\odot}}$, radius $0.96~\rm{R_{\odot}}$, luminosity $0.4~\rm{L_{\odot}}$ and temperature $5040~\rm{K}$), as well as the binary semi-major axis $a=3.3~\rm{R_{\odot}}$, the values used by \citet{2018A&A...615A..81N}. The input parameters $k_{1}$ and $k_{2}$ depend on the stellar structure and were fixed at the values used by \citet{2018A&A...615A..81N}, i.e. 0.063, 2.075 and 0.125, 1.139 for $R_{d}/R= 0.85$ and 0.75, respectively. The Applegate calculator returns the minimally required energy $\Delta E/E$ needed to drive the observed period modulation in V471~Tau as $0.025 \sim 0.042$. This means that the Applegate mechanism is indeed feasible from an energy point of view.  

However, the third component solution has its limits. There are sets of data that do not fit well with the model of a brown dwarf. Therefore, we also investigated a model which simultaneously combines the presence of a third component and the Applegate mechanism in the system. 
From the DCDFT analyses of the timing residuals (see Section~\ref{subsec:thirdcomp}), we found tentative periods, cycles of $\sim1500, 1900, 2700$ and $4000~\rm{days}$ (or $\approx4, 5, 7$ and $11~\rm{years}$, respectively). The appropriate values of $\Delta P/P$ for these cycles are $2.2\times10^{-6}, 1.7\times10^{-6}, 1.2\times10^{-6}$, and $8.2\times10^{-7}$. 
The span of the timing residuals is $\approx90~\rm{seconds}$. The residuals are distributed unequally, with larger values at the minima of the O-C diagram. We set the amplitude of the timing residuals arbitrarily to $45~\rm{seconds}$, which is half of the observed range. We calculated the fraction of energy needed to drive the changes revealed by the DCDFT analysis, with an amplitude of $45~\rm{seconds}$ for the residuals. As input for the Applegate calculator, we used the same values for the mass, radius, and luminosity of the secondary star and the binary semi-major axis. We also used the same $R_{d}/R$ values of 0.85 and 0.75, respectively.
We found the highest $\Delta E/E$ values ($\Delta E/E\sim0.7$) for $R_{d}/R=0.75$ and the shorter cycle of $\approx4~\rm{years}$. The fraction of total energy decreases to 0.034 for the 11 \rm{year} cycle. The range of energy needed for $R_{d}/R=0.85$ was calculated to be 0.41 for a $\approx4~\rm{year}$ modulation and decreased to 0.02 for the $\approx11~\rm{year}$ cycle. Therefore, the magnetically active secondary star needs more energy to produce the observed changes on a shorter timescale. The required energy also decreases with a smaller amplitude of the variations. 

\section{Discussion}
\label{sec:discussion}

V471~Tau is an interesting object by which to study the mechanisms causing periodic fluctuations in observed eclipse timings. Previous papers by \citet{2007AJ....134.1206K, 2015ApJ...800L..24H, 2015ApJ...810..157V, 2018A&A...620A..42V, 2018A&A...615A..81N} and many others have explored and refined the models of Applegate-type mechanisms, third body components and theories of the evolution of third components in common envelope binaries.

In this paper, we present 32~new times of eclipse of the well-studied PCEB V471~Tauri. These data significantly contribute to the discussion about the cause of the observed period variations. \citet{2022arXiv220606919P} recommend observational timeframes covering at least two circumbinary orbits. We have covered the second cycle of the variations, thus allowing improvements to estimates of the modulation period, possible companion parameters or the energetic feasibility of the Applegate mechanism. 

We added newly acquired times of eclipse to the O-C originally constructed by \citet{2011Ap&SS.331..121K}. Their previous ephemeris does not correspond to new data, but the difficulty of predicting future eclipse times in PCEBs with ETVs is known and was discussed in a paper by \citet{2022arXiv220606919P}.
We have constructed a new and improved ephemeris $T_{0}, P_{0}, dP/dt$ (exact values in Table~\ref{tab:v471tau_3t}) of the binary to rectify this discrepancy. Our new ephemeris is in agreement with the ephemeris calculated on a shorter observational timeline as reported by \citet{2015ApJ...810..157V}. This increases confidence in the previously reported solutions for the ephemeris. Our data reaffirm the wave-like structure of the period variations seen in V471~Tau.

We calculated the parameters of a possible third component to the V471~Tau binary. The third body was found to be on an eccentric orbit with $e=0.393$ and a period of $\approx30~\rm{years}$. The orbital inclination of this component to the binary orbital plane is unknown; therefore, we calculated its mass for five typical inclination values in the range $45~\mathrm{deg}-85~\mathrm{deg}$.
Due to the cadence of observations which mainly follows the times of the binary transits, there are no indications of any occultations caused by the potential third body. It is, however, unlikely to have a co-planar orbit to the known binary.
For an inclination of $85~\mathrm{deg}$, the mass of the component is $0.035\,\rm{M_{\odot}}$ and is consistent with a brown dwarf. We also calculated the maximum separation of the third body from the binary and the separation for the date of the SPHERE VLT direct imaging of the object. From our analysis, the separation at the moment of the SPHERE VLT observation is 210~mas from our analysis. The brown dwarf was therefore closer to the detection limit than the predicted 260~mas reported by \citet{2015ApJ...800L..24H}. 
Also from our analysis, the mass of the brown dwarf, $0.035\,\rm{M_{\odot}}$ is less than the value of $0.044\,\rm{M_{\odot}}$ calculated by \citet{2015ApJ...800L..24H}. If our value for the mass of the companion holds, it should be a fainter object and the brown dwarf closer to the detection limit of the SPHERE VLT observations during the SPHERE VLT observing run in December 2014 and could have remained undetected in the data.
We note the mass of $0.035\,\rm{M_{\odot}}$ cited above assumes that the orbit of the brown dwarf is coplanar with the orbit of the binary. For an inclination of $45~\mathrm{deg}$, the mass, brightness and separation would satisfy the condition for the SPHERE VLT detection as described in \citet{2015ApJ...800L..24H}. Unfortunately we do not have any information about direct imaging of the object during 2018-2019, as proposed in \citet{2015ApJ...810..157V}. 

We analysed the residuals left after removing the effect of the third component (see~Fig.~\ref{fig:2021OC}). The timing residuals are not distributed equally above and below the zero level. The residual values around JD~$2\,448\,000$, arising from data published in \citet{1994A&A...281..811I}, were discussed in \citet{2011Ap&SS.331..121K}. They considered the inaccuracies in the eclipse times and we also observed similar unexpected features around JD~$2\,459\,000$ including the new eclipse times, which made us reconsider their cause. \citet{2015ApJ...810..157V} also mentioned similar residuals around JD~$2\,453\,000$. 

The 45~second amplitude of the residuals after removing the effect of the brown dwarf, which is similar to the $40~\rm{seconds}$ reported by \citet{2018RNAAS...2..179M}, motivates us to consider alternative explanations to the third component solution. Our work revealed that although the third body model is possible, it is not entirely satisfactory. 

Despite many works favouring the multi-component scenario, the effect of the Applegate mechanism should be considered. Even with future updates of the third body model parameters, we suspect the third companion model  will not be sufficient to explain the timing residuals with an amplitude of $\approx45~\rm{seconds}$ already seen in the current data with cycles between $4-13$~years.

Therefore, based on the specified period modulation of $\approx30 \rm{years}$ and its $\approx230~\rm{second}$ amplitude, we investigated the amount of energy required to drive the change of the gravitational quadrupole moment of the companion star in such a way that the observed period modulation can be generated by changes in the shape of the companion. We investigated the Applegate mechanism as proposed by \citet{2018A&A...615A..81N} using the online Applegate tool. The refined modulation period and the amplitude of the cyclic changes, together with the parameters describing the magnetically active secondary star (see Section~\ref{subsec:applegatemagneto} and the Applegate online tool) confirm the energetic feasibility of the Applegate mechanism as the driving mechanism for the changes observed in the ETVs of V471~Tau. The required fraction of energy of the  active star $\Delta E/E$ is 0.025 - 0.042 and is about $\approx70$ per cent of the estimates published by \citet{2018A&A...615A..81N}. 

The energetic feasibility of the Applegate mechanism for the secondary star of~V471~Tau presented in this work is in agreement with the results reported for eclipsing binaries with similar components by \citet{2016MNRAS.460.3873B}. The Applegate mechanism can, therefore, explain the eclipse timing variation of V471~Tau. We have to note, that a $\approx30\,\rm{yr}$ activity cycle is not typical for a fast-rotating K~star. In recent study on 120~RS~CVn stars \citet{2022MNRAS.512.4835M} did not found activity cycle longer than 20~yr. However, this could be due to the length of the data used in their work.

For the spin-orbital coupling mechanism presented by \citet{2020MNRAS.491.1820L}, the reported times of eclipse confirm the modulation period of $\approx30\,\rm{years}$ and are consistent with the estimates of the required energy to drive the period variations calculated by \citet{2020MNRAS.491.1820L}.

We also investigated the possibility of the simultaneous presence of a third component in the system and an Applegate mechanism responsible for the cyclic variations seen in the timing residuals. The results of our period analysis of the timing residuals show a wide power peak around the cycle of 11~years and extra peaks for cycles of 4, 5, and 7~years. We found shorter cycles than the values reported by \citet{2015ApJ...810..157V} ($5, 9$ and $13~\rm{years}$).
In recent studies, \citet{2021A&A...650A.158K, 2022MNRAS.513.2893Z} found evidence of a possible activity cycle length of $\approx13~\rm{yr}$, which matches the wide peak seen in the periodogram of the O-C brown dwarf model residuals. Tentative periods found in our period searches are consistent with results from long-term X-ray study by \citet{2022RNAAS...6...94C}.
We set the amplitude of the fluctuation in the timing residuals arbitrarily to 45~seconds and computed that the fraction of the available energy in the secondary star required to drive variations in the timing residuals at the detected cycles is in the range $0.02\sim0.7$. We can explain the variations found in the timing residuals by the Applegate mechanism presented by \citet{2018A&A...615A..81N}. 
However, the amplitude of the fitted sinusoids to the timing residuals published by \citet{2015ApJ...810..157V} is only $\approx6~\rm{seconds}$.
The fraction of the energy needed to cause variations with smaller amplitudes and longer periods, such as those presented by \citet{2015ApJ...810..157V}, is only 0.006.

The missing detection of the third component via direct imaging does not rule out its existence. However, the period analysis of the residuals after modelling the third component revealed the need for an additional quasi-periodic effect such as the proposed Applegate mechanism in the secondary star. We calculated that an Applegate-type mechanism can explain the observed cyclic changes in the timing residuals. Moreover, the Applegate or Lanza-type mechanism can explain the period modulation without the presence of an additional component in the system. \citet{2020MNRAS.491.1043N, 2022arXiv220503163N} numerically simulated quasi-periodic quadrupole moment variations in a solar mass star with 3 and 20~times the solar rotation rate, similar to, but still below the rotation rate of the secondary in V471~Tau. Their 3D magnetohydrodynamic simulations produced variations on timescales similar to those found in our DCDFT analysis, although roughly still two orders of magnitude lower than the 45~second amplitude detected in the timing residuals for the slow rotator model and about a factor 2 smaller for their fast rotator model.

The presence of a third body in the system would cause exact periodical ETVs with residuals caused by the eclipse timing accuracy. Since we observed only the second cycle of ETVs, future eclipse times will reveal the confidence in the presented brown dwarf model. However, the timing residuals indicate additional quasi-periodic variations, which are probably related to the changes of the stellar quadrupole moment as a result of magnetic activity inside the V471~Tau secondary. The significant development of the magnetohydrodynamic models, along with further observations of the times of eclipse, and additional observational studies of PCEBs will get us closer to the combination of processes explaining the ETVs in V471~Tau.

\section{Conclusions}
\label{sec:conclusion}
We present 32~eclipse times of the PCEB V471~Tau falling into the second only cycle of the period modulation observed in this system. We can explain the eclipse timing variation by the presence of a third component or by the Applegate or Lanza-type mechanisms interior to the secondary, magnetically active star, or by combining a third component and the Applegate mechanism in the secondary magnetically active star.
To explain the observations, the third component must be a brown dwarf on an eccentric orbit with a $\approx30~\rm{year}$ period and a mass of $0.035\,\rm{M_{\odot}}$ assuming coplanarity of its orbit with the orbit of the binary. 
From our calculations, the brown dwarf is close to the detection limit of the binary SPHERE VLT direct imaging observations in 2014.
The wave-like distribution of the timing residuals, after removing the effect of the brown dwarf needs, special attention. We found four tentative periods in the timing residuals and investigated the energetic feasibility of the Applegate mechanism. We conclude that a fraction of $0.02\sim0.7$ of the magnetically active star's internal energy can explain the variations in the timing residuals. 

Separately, we investigated the Applegate-type mechanism as the cause of the $\approx30~\rm{year}$ period modulation, i.e. without the presence of an additional component in the system. We found only $0.025\sim 0.042$ of the energy of the magnetically active red dwarf in V471~Tau is needed to drive the change of gravitational quadrupole moment generating the observed eclipse timing variation.

The Applegate-type mechanism, therefore, needs to be included as part of the explanation of the observed period variations in V471~Tau, regardless of the presence of a third component.

\section*{Acknowledgements}

The authors thank the anonymous referee for useful comments that helped to improve this work. The authors thank V.~Koll\'{a}r for his technical assistance. This work was supported by the VEGA grant of the Slovak Academy of Sciences No. 2/0031/22. This work was also supported by the grant APVV-20-0148 ("From Interacting Binaries to Exoplanets") of the Slovak Research and Development Agency.

\section*{Data Availability}

Raw data were generated at the Astronomical Institute of the Slovak Academy of Sciences, Slovakia, at Astrolab IRIS observatory, Belgium and at Vihorlat Observatory, Slovakia. The derived data supporting the findings of this study are available from the corresponding author (EK) upon request. 
 


\bibliographystyle{mnras}
\bibliography{kundra-v471tau-fl} 





\bsp	
\label{lastpage}
\end{document}